\documentclass[letterpaper,twocolumn,english,aps,prb,floatfix,showpacs,amsfonts]{revtex4}
\usepackage[T1]{fontenc}
\usepackage[latin1]{inputenc}
\usepackage{amsmath}
\usepackage{babel}
\usepackage{graphics}
\usepackage{amssymb}

\makeatletter

\usepackage{bm}

\makeatother
\begin{document}

\title{Small Fermi surface in the one-dimensional Kondo lattice model}

\author{J. C. Xavier}

\email{jcxavier@ifi.unicamp.br}

\affiliation{Instituto de Física Gleb Wataghin, Unicamp, Caixa Postal 6165, 13083-970
Campinas, São Paulo, Brazil}

\author{E. Novais}

\email{peres@ifi.unicamp.br}

\affiliation{Instituto de Física Gleb Wataghin, Unicamp, Caixa Postal 6165, 13083-970
Campinas, São Paulo, Brazil}

\author{E. Miranda}

\email{emiranda@ifi.unicamp.br}

\affiliation{Instituto de Física Gleb Wataghin, Unicamp, Caixa Postal 6165, 13083-970
Campinas, São Paulo, Brazil}

\date{\today}

\begin{abstract}
We study the one-dimensional Kondo lattice model through the density
matrix renormalization group (DMRG). Our results for the spin correlation
function indicate the presence of a small Fermi surface in large portions
of the phase diagram, in contrast to some previous studies that used
the same technique. We argue that the discrepancy is due to the open
boundary conditions, which introduce strong charge perturbations that
strongly affect the spin Friedel oscillations.
\end{abstract}


\maketitle
Several uncertainties still exist in our understanding of the physics
of heavy fermion materials.\cite{hewson} The importance of solving
these uncertainties has become even more pressing as we attempt to
understand the anomalous behavior observed in the vicinity of the
clean antiferromagnetic quantum critical point.\cite{pierspepinsirevaz}
The two major scenarios take radically different points of view. In
the first one, conduction electrons are assumed to acquire their peculiar
dynamics through an essentially perturbative coupling to the slow
critical modes of the antiferromagnetic background.\cite{hertz,millis}
Alternatively, the starting point is taken to be the strong coupling
of the conduction electrons and the localized spins to form singlets,
as in the single impurity Kondo problem. The nature of the quantum
critical point is then linked to a non-trivial competition between
the local dynamics and the long wavelength antiferromagnetic fluctuations.\cite{sietal}

One of the key assumptions of the second approach is the presence
of a large Fermi surface (FS), namely one whose volume is given by
including the localized spins in the count\[
\frac{2V^{*}_{FS}}{\left( 2\pi \right) ^{D}}=n+1.\]
Behind this assumption lies a deep connection\cite{martin} between the
Friedel sum rule of the single impurity Kondo problem\cite{friedel}
and Luttinger's theorem for the FS volume of a system of interacting
fermions.\cite{luttingertheorem} The inclusion of the localized
electron in the count is natural within an Anderson lattice model
description at weak coupling but becomes doubtful at strong coupling
where the Kondo lattice Hamiltonian is the effective low-energy
theory. However, topological arguments have been used, in
one\cite{yamanakaetal} as well as in higher\cite{oshikawa} dimensions,
to show that indeed neutral gapless excitations must exist at a
\textbf{\( \mathbf{k} \)}-vector spanning a large FS.%
\footnote{We adhere to the somewhat inappropriate yet common usage of the term
Fermi {}``surface'' in one dimension.
} Furthermore, numerical studies of the one-dimensional Kondo lattice
model have also pointed to the presence of a large FS.\cite{shibataetal, shibataetal1}
In this paper, we show that a more careful analysis of the numerical
evidence casts serious doubts on these conclusions and leaves open
the question of the size of the Fermi surface in heavy fermion systems.

We considered the one-dimensional \( S=\frac{1}{2} \) Kondo lattice
Hamiltonian with \( L \) sites \[
H=-t\sum ^{L-1}_{j=1,\sigma }c^{\dagger }_{j,\sigma }c^{\phantom {\dagger }}_{j+1,\sigma }+J\sum ^{L}_{j=1}\mathbf{S}_{j}\cdot \mathbf{s}_{j}\]
 where \( c_{j\sigma } \) annihilates a conduction electron in site
\( j \) with spin projection \( \sigma  \), \( \mathbf{S}_{j} \)
is a localized spin \( \frac{1}{2} \) operator and \( \mathbf{s}_{j}=\frac{1}{2}\sum _{\alpha \beta }c^{\dagger }_{j,\alpha }\bm {\sigma }_{\alpha \beta }c^{\phantom {\dagger }}_{j,\beta } \)
is the conduction electron spin density operator. \( J>0 \) is the
Kondo coupling constant between the conduction electrons and the local
moments and the hopping amplitude \( t \) is set to unity to fix
the energy scale. We treated the model with the density matrix renormalization
group (DMRG) technique\cite{white, white2} with open boundary conditions.
We used the finite-size algorithm for sizes up to \( L=120 \) keeping
up to \( m=400 \) states per block. The discarded weight was typically
about \( 10^{-5}-10^{-8} \) in the final sweep.

There is compelling evidence\cite{Tsunetsugu} that the one-dimensional
Kondo lattice model away from half-filling is a Tomonaga-Luttinger
(TL) liquid.\cite{voit} TL liquids with periodic boundary conditions,
have charge and spin correlation functions given asymptotically by\begin{eqnarray}
\left\langle \delta n\left( 0\right) \delta n\left( x\right) \right\rangle  & = & \frac{K_{\rho }}{\left( \pi x\right) ^{2}}+A_{1}\frac{\cos \left( 2k_{F}x\right) }{x^{K_{\rho }+1}}\nonumber \\
 & + & A_{2}\frac{\cos \left( 4k_{F}x\right) }{x^{4K_{\rho }}},\label{nn-corr} \\
\left\langle \mathbf{S}\left( 0\right) \cdot \mathbf{S}\left( x\right) \right\rangle  & = & \frac{1}{\left( \pi x\right) ^{2}}+B_{1}\frac{\cos \left( 2k_{F}x\right) }{x^{K_{\rho }+1}},\label{ss-corr} 
\end{eqnarray}
 where \( K_{\rho } \) is a non-universal correlation exponent and
\( k_{F} \) is the Fermi momentum. Besides, local charge and spin
perturbations, such as introduced by impurities or boundaries, lead
to corresponding Friedel oscillations, which in the case of a TL liquid
take the form\cite{shibataetal,shibataetal1,eggertaffleck,eggergrabert,voitwanggrioni}

\begin{eqnarray*}
\left\langle \delta n\left( x\right) \right\rangle  & = & C_{1}\frac{\cos \left( 2k_{F}x\right) }{x^{\left( K_{\rho }+1\right) /2}}+C_{2}\frac{\cos \left( 4k_{F}x\right) }{x^{2K_{\rho }}},\\
\left\langle \delta S_{z}\left( x\right) \right\rangle  & = & D_{1}\frac{\cos \left( 2k_{F}x\right) }{x^{K_{\rho }}}.
\end{eqnarray*}

\begin{figure}
{\centering \resizebox*{3.1in}{!}{\includegraphics{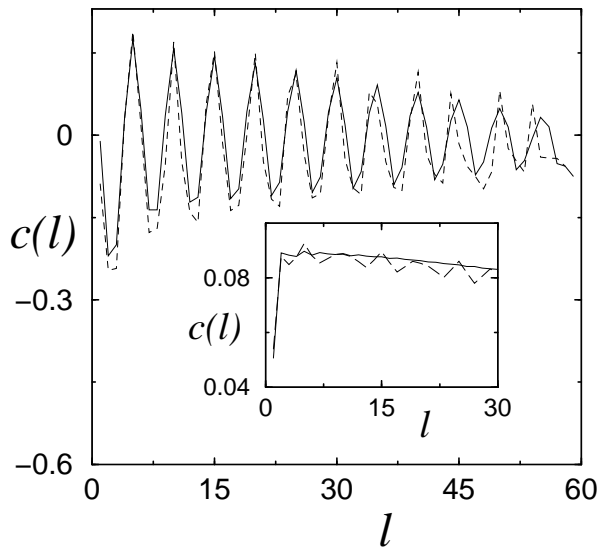}} \par}

\caption{\label{fig1} Spin-spin correlation function \protect\( c\left( l\right) \protect \)
for \protect\( n=\frac{2}{5}\protect \), \protect\( L=60\protect \)
and \protect\( J=0.35\protect \) (main) and \protect\( J=5\protect \)
(inset). The solid lines correspond to lattice averages, whereas the
dashed lines are obtained with the two sites equidistant from the
chain center (see text).}
\end{figure}

\begin{figure}
{\centering \resizebox*{3.1in}{!}{\includegraphics{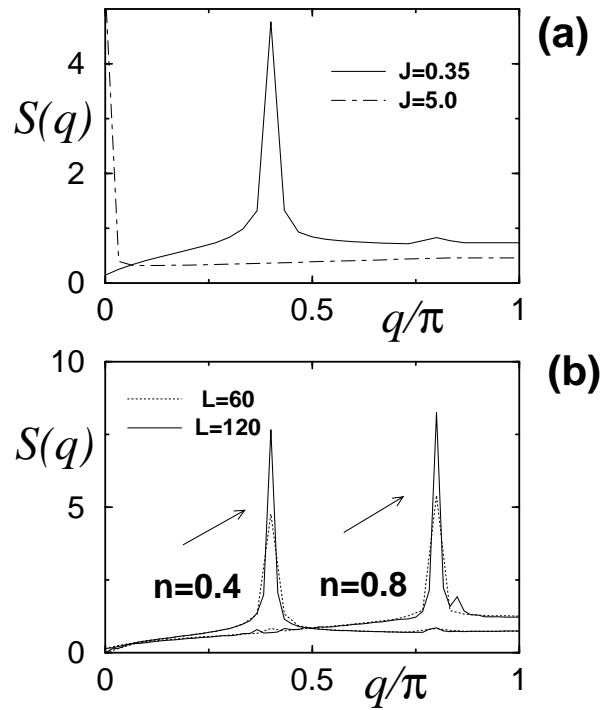}} \par}

\caption{\label{fig2} Spin structure factor S(q): (a) \protect\( n=\frac{2}{5}\protect \),
same parameters as in Fig.~\ref{fig1}; (b) \protect\( n=\frac{2}{5}\protect \)
and \protect\( n=\frac{4}{5}\protect \) for both \protect\( L=60\protect \)
and \protect\( L=120\protect \) (all at \protect\( J=0.35\protect \)).}
\end{figure}

We first show our results for the total spin correlation function
\( c\left( j,k\right) =\left\langle \mathbf{S}^{T}_{j}\cdot \mathbf{S}^{T}_{k}\right\rangle  \),
where \( \mathbf{S}^{T}_{j}=\mathbf{S}_{j}+\mathbf{s}_{j} \). Since
we use open boundary conditions, translational invariance is lost
and \( c\left( j,k\right)  \) depends on both \( j \) and \( k \).
We present results obtained by two different methods. In the first
one (dashed lines in Fig.~\ref{fig1}), \( c\left( l=\left| j-k\right| \right)  \)
was obtained by taking \( j \) and \( k \) at the same distance
(within a lattice spacing) from the center of the chain. We call this
the central value of \( c\left( l\right)  \). In the second one (solid
lines in Fig.~\ref{fig1}), we averaged over all pairs of sites
separated by the distance \( l=\left| j-k\right|  \). We will call
it the average value of \( c\left( l\right)  \). If the boundary
perturbation has a negligible effect on the spin-spin correlation
function, then the two methods should yield similar results and we
can have confidence that we have obtained the bulk value of \( c\left( l\right)  \).
This is indeed what is observed for the density \( n=\frac{2}{5} \)
with \( L=60 \), as seen in Fig.~\ref{fig1}. The paramagnetic
(PM) and ferromagnetic (FM) phases\cite{Tsunetsugu} can be easily
discerned from the long distance behavior of \( c\left( l\right)  \):
for \( J<J_{c} \), the envelope of \( c\left( l\right)  \) tends
to zero (main plots in Fig.~\ref{fig1}) and for \( J>J_{c} \)
it tends to the magnetization squared (inset of Fig.~\ref{fig1}).
The critical value is \( J_{c}\approx 1.2 \) for \( n=\frac{2}{5} \).
The ferromagnetism was also checked directly from the total spin of
the ground state. The PM phase exhibits well developed Friedel oscillations
due to the open boundaries.

\begin{figure}
{\centering \resizebox*{3in}{!}{\includegraphics{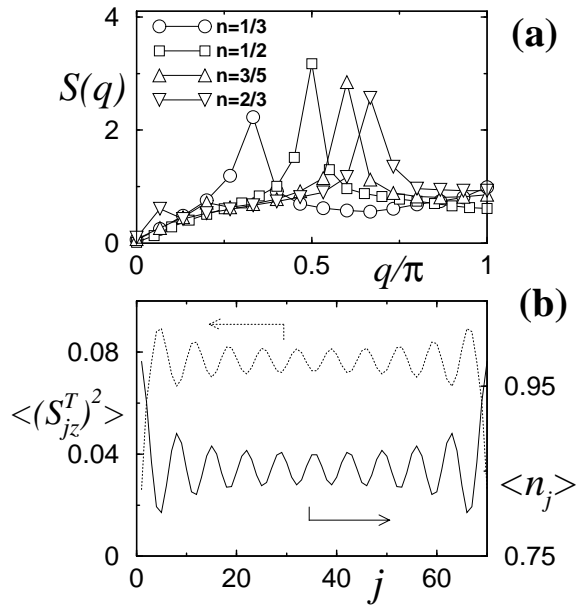}} \par}

\caption{\label{fig3} (a) \protect\( S(q)\protect \) vs. momentum for several
densities. In all cases, \protect\( J=0.5\protect \) and \protect\( L=30\protect \),
except for \protect\( n=\frac{1}{2}\protect \), where \protect\( L=40\protect \).
(b) \protect\protect\( \left\langle \left( S^{T}_{jz}\right) ^{2}\right\rangle \protect \protect \)
and \protect\( \left\langle n_{j}\right\rangle \protect \) vs. distance
for \protect\( L=70\protect \), \protect\( J=2.5\protect \), and
\protect\( n=\frac{6}{7}\protect \). Peaks of \protect\protect\( \left\langle \left( S^{T}_{jz}\right) ^{2}\right\rangle \protect \protect \)
track the valleys of \protect\( \left\langle n_{j}\right\rangle \protect \).}
\end{figure}

In Fig.~\ref{fig2}(a), we show the spin structure factor \( S\left(
q\right) =\frac{1}{L}\sum _{j,k}e^{iq\left( j-k\right) }c\left(
j,k\right)  \) 
corresponding to Fig.~\ref{fig1}. Because of the weak influence
of the boundaries, this is very close to the Fourier transform of
\( c\left( l\right)  \). Whereas \( S\left( q\right)  \) is maximum
at \( q=0 \) in the FM phase (with \( LS\left( 0\right) =S^{T}\left( S^{T}+1\right)  \)),
in the PM phase \( S\left( q\right)  \) is peaked at \( q_{s}=\pi n \)
at \( n=\frac{2}{5} \).%
\footnote{The residual value of \( S\left( q=0\right)  \) in the PM phase is
due to the DMRG truncation at \( m=400 \). We have checked that it
decays exponentially to zero as \( m \) is increased.  } This result
does not seem to be due to finite size effects. Indeed, the structure
factor peak gets sharper and more pronounced as one goes from \( L=60
\) to \( L=120. \) This is shown in Fig.~\ref{fig2}(b), for the two
densities \( n=\frac{2}{5} \) and \( n=\frac{4}{5} \) at \( J=0.35. \)
This is \emph{strong evidence for a small Fermi surface} with \(
2k_{F}=q_{s}=\pi n \), in sharp contrast to what was previously
reported.\cite{shibataetal,shibataetal1} Note a very small feature at
wave vector $2\pi n$, which, however, does not increase with the
system size and is below the accuracy of the DMRG.  We also calculated
the spin structure factor at several other density values, as shown in
Fig.~\ref{fig3}(a). As finite size effects appear to be negligible, we
have kept to smaller chain sizes (\( L=40 \) or \( 30 \)). In all
cases, there was good agreement between central and average values of
\( c\left( l\right) \). All the results point to the presence of a
small FS. In order to understand the origin of this discrepancy we
re-examined the parameter ranges studied in
Refs.~\onlinecite{shibataetal} and \onlinecite{shibataetal1}. As we
will see, their results occur at large values of \( J \) (\( J\agt 1
\)), where strong boundary charge perturbations mask the true bulk
behavior of spin correlations.  This is diagnosed by very different
values of the average and the central \( c\left( l\right) \). On the
other hand, when \( J\alt 1 \) as in all cases seen above, the
boundaries induce only a weak charge disturbance. As a result, central
and average \( c\left( l\right) \) are the same, and the spin
correlation function oscillates at \( 2k_{F}=q_{s}=\pi n \).

\begin{figure}
{\centering \resizebox*{3in}{!}{\includegraphics{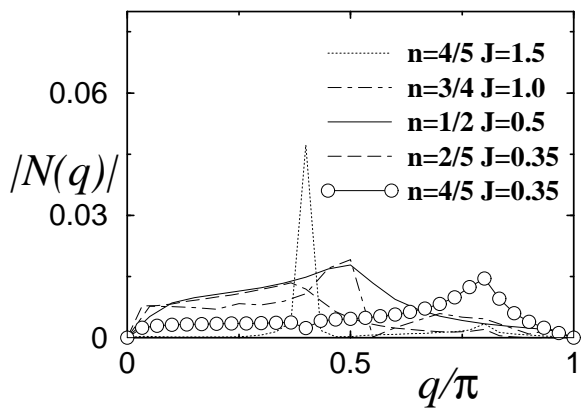}} \par}

\caption{\label{fig4} \protect\( \left| N\left( q\right) \right| \protect \) vs. momentum
for several densities. Coupling constants are indicated and lattice
sizes are, from top to bottom, \protect\( L=60,40,40,60,\protect \)
and 60, respectively.}
\end{figure}

\begin{figure}
{\centering \resizebox*{3.1in}{!}{\includegraphics{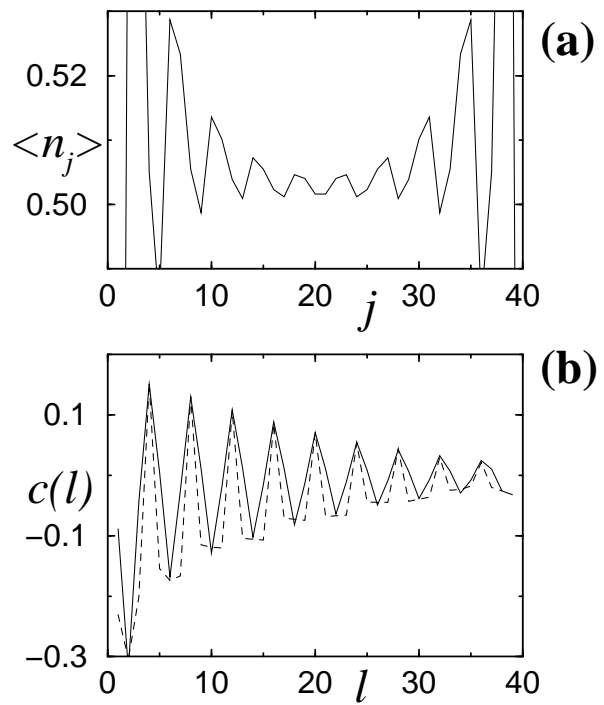}} \par}

\caption{\label{fig5} (a) \protect\( \left\langle n_{j}\right\rangle \protect \)
vs. distance for \protect\( L=40\protect \), \protect\( J=0.5\protect \)
and \protect\( n=\frac{1}{2}\protect \). (b) Spin-spin correlation
function for \protect\( n=\frac{1}{2}\protect \) with \protect\(
J=0.5\protect \) and \protect\( L=40\protect \). Solid  and dashed
lines correspond to average and central $c(l)$, as in Fig.~\ref{fig1}.}
\end{figure}

In Fig.~\ref{fig3}(b), we present the local density \( \left\langle
n_{j}\right\rangle \) and the squared \( z \)-component of the total
spin \( \left\langle \left( S^{T}_{jz}\right) ^{2}\right\rangle \)
versus distance for the density \( n=\frac{6}{7} \) and \( J=2.5 \).
The charge Friedel oscillations induced by the open boundaries are the
same as found previously\cite{shibataetal, shibataetal1} and the
squared \( z \)-component is anti-correlated with the charge.  This
strong charge disturbance is seen at other densities when \( J\agt 1
\) but is suppressed as \( J \) is decreased. This can be seen in
Fig.~\ref{fig4}, where we show the magnitude of the Fourier transform
\( N\left( q\right) =\frac{1}{L}\sum _{j}e^{iqj}\left( \left\langle
n_{j}\right\rangle -n\right) \) versus momentum for several densities
and coupling constants. Furthermore, when \( J \) is decreased, the
charge Friedel oscillation peak moves from \( q_{c}=2\pi \left(
1-n\right) \) to \( q_{c}=2k_{F}=\pi n \).  We note that a peak at \(
q_{c}=2\pi \left( 1-n\right) \) (mod \( 2\pi \)) cannot distinguish
between a small FS, with \( 4k_{F}=2\pi n \), and a large one, with \(
4k_{F}^{*}=2\pi \left( 1+n\right) \) (we denote a large Fermi vector
by a star). For this reason, we cannot use the charge structure factor
to determine the size of the Fermi surface.  Even when the peaks are
not sharp, the oscillations are still quite well defined, as shown in
Fig.~\ref{fig5}(a) for \( n=\frac{1}{2} \) (compare the scales of
Fig.~\ref{fig5}(a) and Fig.~\ref{fig3}(b)).

\begin{figure}
{\centering \resizebox*{3.1in}{!}{\includegraphics{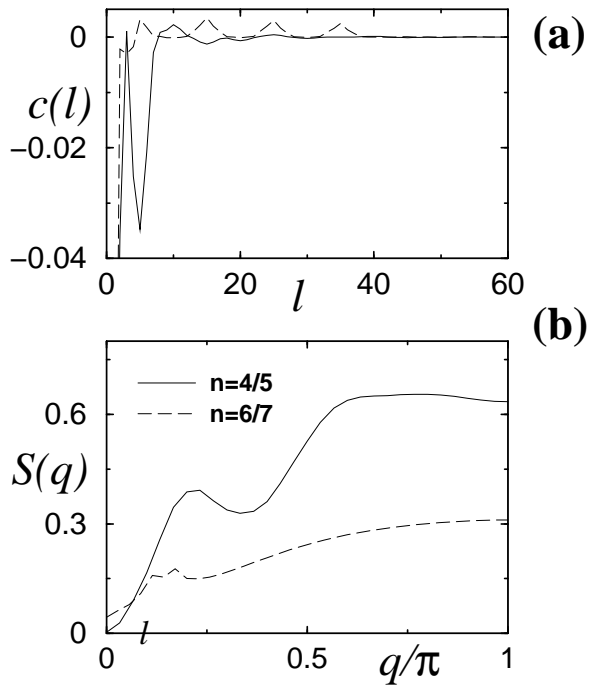}} \par}

\caption{\label{fig6} (a) Average (solid) and central (dashed) $c(l)$
for \protect\( n=\frac{4}{5}\protect \) with \protect\( J=1.5\protect
\) and \protect\( L=60\protect \), as in Fig.~\ref{fig1}. (b)
\protect\( S\left( q\right) \protect \) vs. momentum. The densities
are indicated. The parameters are the same 
as in Figs.~\ref{fig3} and \ref{fig4} for these densities. }
\end{figure}

The presence of strong boundary charge disturbances leads to different
behaviors of the central and the average spin correlation functions.
This connection is made apparent in Figs.~\ref{fig5}(b) and \ref{fig6}(a),
which show average and central \( c\left( l\right)  \) for \( n=\frac{1}{2} \),
\( J=0.5 \) and \( n=\frac{4}{5} \), \( J=1.5 \), respectively.
At \( n=\frac{1}{2} \), \( J=0.5 \), charge oscillations are small
and the average and central \( c\left( l\right)  \) are almost the
same. On the other hand, the large charge peak that occurs at \( n=\frac{4}{5} \),
\( J=1.5 \) (Fig.~\ref{fig4}) leads to quite different values
of the average and central \( c\left( l\right)  \). This is also
reflected in the spin structure factor, which shows only broad ill-defined
features, as plotted in Fig.~\ref{fig6}(b). This is the key to understanding
the discrepancy between our results and the ones of Shibata \emph{et
al.}\cite{shibataetal, shibataetal1}

In order to observe spin oscillations, Shibata \textit{et. al} applied
a small local field at the chain ends and measured \( \left\langle S^{T}_{jz}\right\rangle  \).
We obtained similar spin oscillations for \( \left\langle \left( S^{T}_{jz}\right) ^{2}\right\rangle  \),
but with half the period, \emph{without any field}. The origin of
the latter structure is clear: local spin fluctuations are determined
by the charge Friedel oscillations (see Fig.~\ref{fig3}(b)). Furthermore,
this also shows that the system is rendered more polarizable by the
boundary perturbation at the peaks of \( \left\langle \left( S^{T}_{jz}\right) ^{2}\right\rangle  \),
accounting for the oscillations of \( \left\langle S^{T}_{jz}\right\rangle  \)
when a magnetic field is applied at the endpoints. Thus, the response
of the system under the application of boundary magnetic fields, as
was done in Refs.~\onlinecite{shibataetal} and \onlinecite{shibataetal1},
is strongly perturbed by the presence of open boundaries and cannot
by itself be used to determine the size of the FS. In that case, the
spin oscillations are inextricably linked to the charge ones by \( q_{s}=q_{c}/2 \).
Additional confidence in this picture can be gained by the inspection
of Figs.~2 and 4 of Ref.~\onlinecite{shibataetal}, where an increase
of the charge peak is accompanied by an increase of the spin peak.
Spin Friedel oscillations should ideally be studied by applying a
small local magnetic perturbation \emph{in the absence of any charge
perturbation}. We conclude that, based on the available evidence,
it is impossible to determine whether the system has a large or a
small Fermi surface for \( J\agt 1 \).

We can envisage three alternatives to try to overcome this difficulty.
The first one would be to use larger boundary fields so that the spin
perturbation can surpass the charge one. However, it is not clear
that this regime is attainable without drastically changing the nature
of the ground state. A second way would be to work with periodic boundary
conditions and a magnetic field applied at one site only, thus eliminating
boundary charge perturbations while keeping spin ones. However, this
is not particularly appealing because the DMRG loses accuracy with
periodic boundary conditions. Finally, by going to larger systems
one can have access to the true bulk behavior. Without these further
studies, the size of the Fermi surface at \( J\agt 1 \) remains undetermined.
\begin{figure}
{\centering \resizebox*{3.1in}{!}{\includegraphics{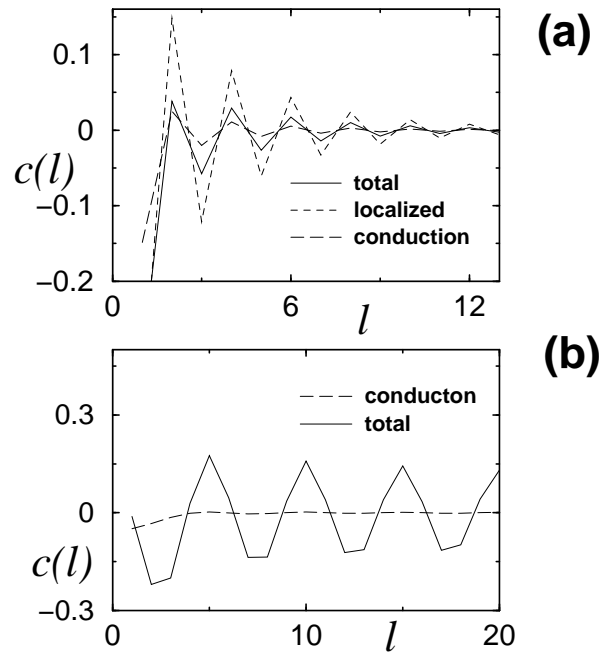}} \par}

\caption{\label{fig7} (a) The average spin-spin correlation function vs.
distance for the total, localized and the conduction electron spins
in a chain of \protect\( L=42\protect \) sites, with \protect\( J=1.0\protect \)
at half filling. (b) Same as (a) but for a chain of \protect\( L=60\protect \)
sites, \protect\( J=0.35\protect \), and \protect\( n=\frac{2}{5}\protect \).
Here, the correlations between localized spins are not shown. Both
in (a) and (b), only the first few sites are shown.}
\end{figure}

We have also calculated the spin correlations between localized spins
only and conduction electrons only. In Fig.~\ref{fig7}(a) and (b)
we present these correlations together with the total spin-spin correlations
for densities \( n=1 \) and \( n=\frac{1}{4} \), respectively. In
Fig.~\ref{fig7}(b), the correlations between localized spins are
not shown since they differ very little from the total spin ones.
As expected all three functions have the same period, differing only
in amplitude. It is also clear that the conduction electron contribution
increases with the density. Note also that the spin correlations at
half filling (Fig.~\ref{fig7}(a)) decay much faster than at other
fillings, due the presence of a spin gap.\cite{tsunetsuguetal,yuwhite}

Let us now discuss these results. The theorem of Ref.~\onlinecite{yamanakaetal}
guarantees that, provided there is either unbroken time reversal or
parity symmetries in the ground state, the one-dimensional Kondo lattice
has gapless excitations at \( 2k_{F}^{*}=\pi \left( 1-n\right) . \)
This should be valid within the PM phase. The conventional Luttinger
liquid phenomenology then tells us that these are collective spin
and/or charge bosonic excitations with a linear dispersion with respect
to deviations from this wave vector. They lead to the characteristic
oscillatory behavior of Eqs.~(\ref{nn-corr}) and (\ref{ss-corr})
and the corresponding peaks in the spin and charge structure factors.
Of course, the theorem does not forbid the appearance of gapless excitations
at \emph{other} wave vectors such as \( 2k_{F}=\pi n \). Together
with our results, this would seem to indicate that the spectral weight
at \( 2k_{F}^{*} \) is rather small in the range \( J\alt 1 \),
most of it being concentrated at \( 2k_{F} \). 

The conventional wisdom about heavy fermion systems is based on the
competition between the local Kondo effect and the inter-site
Ruderman-Kittel-Kasuya-Yosida (RKKY) interaction. Although this
dichotomy is controversial in one dimension, it is tempting to use it
as a general guide. The size of the compensating Kondo cloud around a
single localized moment has been argued to be exponentially large \(
a\, e^{1/\rho J} \), where \( a \) is the lattice spacing and \( \rho
\) is the density of states at the Fermi level,\cite{sorensenaffleck}
with typical values on the order of 1 \( \mu \)m. Thus, it would
increase as \( J \) is decreased towards the physically relevant
region \( J\alt 1 \), where we observe the peak at \( q_{s}=2k_{F}=\pi
n \). This might lead us to think that we should work with system
sizes that are at least as large as the Kondo compensating cloud in
order to observe features characteristic of a large FS. However, the
fact that the peak at \( 2k_{F} \) becomes
\emph{sharper and more pronounced} as \( L \) is increased (Fig.~\ref{fig3}(b))
casts doubt on this naive expectation. Moreover, even if the true
thermodynamic limit of the spin correlations do indeed oscillate at
\( 2k_{F}^{*}=\pi \left( 1+n\right)  \), our results show that perhaps
\emph{at the physically relevant distance scale} the important wave
vector is actually \( 2k_{F}=\pi n \). For example, neutron scattering
experiments are limited by the coherence length of the neutron beam
and may not be able to probe large distances such as \( a\, e^{1/\rho J} \).
Other probes of the FS size, such as quantum oscillations, are limited
by the electron mean free path, which would also have to exceed this
length scale to access the asymptotic limit. Thus, our results put
stringent limits on the observability of a large FS in heavy fermion
systems. Besides, the presence of disorder and/or inelastic scattering
may render the small FS size the only relevant length scale in real
systems.

In conclusion, we have found clear signatures of a small Fermi surface
in the spin correlation function of the one-dimensional Kondo lattice
at small values of the Kondo coupling constant (\( J\alt 1 \)). The
discrepancy with previous results in the literature that had argued
for a large Fermi surface in this system is ascribed to the presence
of large edge perturbations introduced by the use of open boundary
conditions. These perturbations are larger for large coupling constants
(\( J\agt 1 \)), which hinders the investigation of the Fermi surface
size in this region. Even if the true asymptotic spin correlations
are peaked at the large Fermi surface wave vector, the relevant oscillation
period in many cases of interest may be set by the size of the small
Fermi surface.

JCX is grateful to A. L. Malvezzi for providing some DMRG data for
comparison. We also thank I. Affleck for constructive criticism and
suggestions. This work was supported by the Brazilian agencies FAPESP
(00/02802-7 and 01/00719-8) and CNPq (301222/97-5).

\end{document}